\documentclass[12pt,epsf]{article}
\usepackage{epsfig}

\newcommand{\onefigure}[2]{\begin{figure}[htbp]
\begin{center}\leavevmode\epsfbox{#1.eps}\end{center}\caption{#2\label{#1}}
\end{figure}}

\newcommand{\twofigures}[3]{\begin{figure}[htdp]
\centering \leavevmode\epsfxsize=2.5in\epsfbox{#1.eps}
\leavevmode\epsfxsize=2.5in\epsfbox{#2.eps} 
\caption{{
#3}\label{#1}}
\end{figure}}

\newcommand{\fourfigures}[5]{\begin{figure}[htdp]
\centering \leavevmode\epsfxsize=2in\epsfbox{#1.eps}
\leavevmode\epsfxsize=2in\epsfbox{#2.eps}
\leavevmode\epsfxsize=2in\epsfbox{#3.eps} 
\leavevmode\epsfxsize=2in\epsfbox{#4.eps} 
\caption{{
#5}\label{#1}}
\end{figure}}

\renewcommand{\thanks}[1]{\footnote{#1}} 

\newcommand{\be}{\begin{equation}}
\newcommand{\ee}{\end{equation}}
\newcommand{\bea}{\begin{eqnarray}}
\newcommand{\eea}{\end{eqnarray}}

\begin{document}

\bigskip\bigskip
\begin{center}
{\bf \large Global Causal Structure of a Transient Black Object}
\end{center}

\begin{center}
Tehani Finch \footnote{e-mail address, tkfinch@howard.edu}  and
James Lindesay\footnote{e-mail address, jlindesay@fac.howard.edu}\\
Computational Physics Laboratory \\
Howard University,
Washington, D.C. 20059 
\end{center}
\bigskip

\begin{center}
{\bf Abstract}
\end{center}

A singularity-free and spherically symmetric
transient black object whose center remains always timelike,
yet directly manifests a trapped region,
has been constructed and numerically implemented.   
The exterior geometry is shown to be similar to that of a long-lived transient
black hole, with a few subtle differences.
The large-scale global structure of the geometry is examined through the
construction of a conformal diagram, which 
exhibits no event horizon and bears resemblance
to that of a Minkowski spacetime. 
Since there is no singularity within the geometry, the evolution of
the exchange of information between timelike observers,
including those that fall through the trapped region, can be directly explored. 
The dynamics of generic ``standard" communications, as well as entangled
communications, is exhibited through both $t$ $vs.$ $r$ and
conformal spacetime diagrams.

\bigskip \bigskip \bigskip

\setcounter{equation}{0}
\section{Introduction}
\indent

Quantum aspects of gravitating systems continue to be actively
explored in the physics literature. 
Cosmologies with trapping surfaces are of particular interest,
since the geometry presents regions of interplay between
quantum and gravitational phenomena.  
In particular, black holes manifest horizons,  which
are lightlike surfaces bounding regions of information exchange.  
For black holes, the very coupling of geometrodynamics to quantum processes
likely implies that descriptions of the spacetime
should qualitatively differ from classical static systems.  
For instance, any temperature associated with a
static black hole generates radiations that modify
the black hole. 
Furthermore, the classical, static horizon
of Schwarzschild geometry is
a $t=\infty$ surface, implying that those exterior to the horizon can never observe
infalling objects reach it. 
However, infalling energies likewise modify
the dynamic surfaces through which they fall.
Therefore, geometries with explicit time dependence are needed to model
dynamic spacetimes
consistent with actual systems.

To gain further insights into the surfaces generated in dynamic
cosmologies, dynamic black holes embedded in
asymptotically Minkowski spacetimes have been		
developed and explored \cite{JLMay07, JLPS2010}. 
The metric forms utilize non-orthogonal coordinates
inspired by river models of stationary spacetimes \cite{rivermodel},
with a temporal dependency parameterized
by the river time rather than the Schwarzschild time. 
The geometries are free of singularities away from $r=0$.
Fixed temporal coordinate curves remain spacelike
surfaces throughout the dynamic geometries, corresponding
to the times measured by certain inertial observers. 
The motivation for developing these dynamic descriptions
is that such a temporal foliation simplifies one's descriptions of
quantum coherent processes on the geometry. 
Those inertial observers satisfying
$u_{obs}^{ct}=1$ have been referred to as
being \emph{geometrically stationary} \cite{BHQG},
and in Robertson-Walker cosmology
such observers are usually
called \emph{co-moving} observers. 
These qualitative differences from Schwarzschild time make
examinations of quantum behaviors and
information dynamics
on such geometries more straightforward.

The results presented will develop as follows. 
In Section \ref{sec:TransientBO}, the geometry of 
an example spherically symmetric transient black object is developed.  Its
stress-energy densities will depend upon time and radial coordinates, and will
satisfy all energy conditions everywhere during accretion, and in the exterior
during evaporation.  Consistency conditions on the geometry are demonstrated,
and lightlike trajectories within and near the trapped region are explored.
Section \ref{sec:BOPenrose} develops the conformal diagram of the
dynamic spacetime, 
demonstrating global causal properties of the
singularity-free and horizon-free geometry.
To complete the discussion, 
Section \ref{DynamicInfo} then examines the recovery of
information temporarily trapped within the transient
black object.  The temporal and kinematic dynamics of the outgoing
communications of an infalling emitter are displayed via
spacetime diagrams and plots of emission rates.  The evolution
of an entangled photon pair, one of which traverses the trapped region,
is also explicitly demonstrated.

\setcounter{equation}{0}
\section{Transient Black Objects \label{sec:TransientBO}}
\indent

Because of their relative mathematical simplicity,
static black holes have been of considerable interest in the study
of quantum mechanics in gravity, as well as their influence upon
the formative dynamics of galaxies.  An uncharged,
spherically symmetric black hole has a spacelike ``center"
$r=0$ that implies the existence of a horizon.  This horizon
delineates the outermost boundary of
a region of spacetime within which all future-trending
trajectories ultimately hit the ``center." 
However, it is possible to construct a geometry
that behaves very similarly to a transient black hole in the exterior,
but has an innermost boundary to the trapped region, for which all causal trajectories have
decreasing radial coordinate \cite{Hayward05, Hossenfelder09}.  
The innermost region then
serves as a ``depository" that temporarily stores any information that
falls into the region of trapped trajectories.  The development of such a
transient black object will be the subject of this section.

\subsection{Form of the metric \label{subsec:metric}}
\indent

The radially dynamic
spacetime metric for a  spherically
symmetric, temporally transient black object
will be developed from the general metric
\be
ds^2 ~=~ -\left ( 1 - {R_M (ct,r) \over r}  \right ) \, (dct)^2 + 
2 \sqrt{R_M (ct,r) \over r} \, dct \, dr + dr^2 + r^2 d\theta^2 +
r^2 sin^2 \theta \, d\phi^2 ,
\label{BOmetric}
\ee
whose properties and virtues have been discussed in \cite{JLMay07, Nielsen06}. 
In this equation, a finite radial mass scale
$R_M (ct,r) \equiv 2 G_N M(ct,r) / c^2$ is a length scale of the
mass-energy content of the black object,
${G^0}_0~ = ~ {1 \over r^2} {\partial \over \partial r}R_M (ct,r)=
-{8 \pi G_N \over c^4} {T^0}_0$.    
The metric takes the
form of Minkowski spacetime
both asymptotically ($r >> R_M $) as well as when the
radial mass scale vanishes. 

Radial trajectories for test particles of mass $m$
in this geometry have 4-velocity
components that satisfy\be
u^r =  - \sqrt{R_M \over r}
u^{ct} \pm \sqrt{(u^{ct})^2 - \Theta_m}
\quad , \quad \Theta_m \equiv \left \{
\begin{array}{l}
1 \quad m \not= 0 \\
0 \quad m = 0
\end{array}  \right .
, \label{4velocities}
\ee
where the + sign signifies ``outgoing" trajectories, and the
- sign signifies ``ingoing" trajectories. 
For massive systems whose proper time
(up to an additive constant) is given
by $t$, the temporal component of their 4-velocities
satisfy $u^{ct}=1$, and the trajectory is
neither ingoing nor outgoing. Freely falling trajectories
sharing this temporal coordinate represent what
have been referred to as \emph{geometrically stationary}
trajectories \cite{BHQG}.  For this physical setting,
observer trajectories with 4-velocity components satisfying
$u_{obs}^{ct}=1, \, u_{obs}^r=  - \sqrt{R_M \over r_{obs}}$
can be shown to satisfy geodesic equations for massive gravitating
systems which share proper time with the asymptotic observer,
$dt=d\tau$.  These are the co-moving observers
of this geometry.

The radial coordinate provides
the length scale for local angular
proper distances $d \ell_\theta = r \, d\theta$
and transverse areas $d^2 \sigma = r^2 sin \theta \, d\theta \, d\phi$. 
One should also note that any geometrically stationary observer
($u^{ct}=1$) in this geometry will measure a proper
radial distance interval at a fixed time value
(i.e. a synchronous
proper length measurement shared by other geometrically
stationary observers) given by $ds=dr$.  This implies that $r$ can also be interpreted as the proper
distance between a geometrically stationary observer with coordinates
$(ct,r)$ and that geometrically stationary observer that
is encountering the
center $r=0$ at the same value of $t$.  This fact motivates the use of the term ``center" for $r=0$  \cite{JLTF2010}.  Such an interpretation does not hold for
fiducial (fixed $r$) observers, who must undergo
accelerations in order to maintain their radial coordinate. 

As can be seen from (\ref{BOmetric}) and
 (\ref{4velocities}),
at the \emph{trapping surfaces} $R_{TS}$, instantaneously given by 
solutions to 
\be
1=\sqrt{ {R_M (ct,R_{TS}) \over R_{TS}} }  ,
\label{TrappingSurfaceEqn}
\ee
 outgoing light will be momentarily
stationary.  
In addition, curves of fixed radial coordinate
labeled by $r$  
are spacelike in the region between these surfaces. 
Thus, if solutions to this equation exist, the surfaces $R_{TS}$ bound
regions that exclude the possibility of having fiducial
observers, since even light cannot be stationary
within these regions.  Such
regions are referred to as \emph{trapped regions} of the
spacetime.  If trapped regions exist, the geometry contains a ``black object."  
If there is a horizon, the geometry contains a ``black hole." 

Several points of interest directly follow from these equations
and the form of the metric:
\begin{itemize}
\item The radial coordinate is a proper distance
as well as a measure of transverse areas for a class of observers;
\item Outgoing photons at the trapping surface are momentarily
stationary in the radial coordinate.  For dynamic geometries, outgoing photons that are
crossed by the outermost trapping surface have 
$u_{\gamma +}^r =0$ at that instant.  Thus, there can be no observers with
stationary or increasing radial coordinate in regions for which
$u_{\gamma +}^r  \le 0$.  The radial scales $R_{TS}$ represent
\textit{static limits} in this geometry. 
These surfaces are sometimes referred to as ``apparent horizons".
\end{itemize}
A calculation of curvature components for the metric (\ref{BOmetric})
exhibits no inherent singularities away from $r=0$.
This means that no observer measures singular curvatures on any
surface or transition time of the geometry. 
The functional form of the radial
mass scale $R_M (ct,r)$ can also be chosen to preclude
any singular behavior at $r=0$ itself.  Such geometries
are thus singularity free. 

It is quite straightforward to choose a form for the
radial mass scale that generates a region in the spacetime
for which the center $r=0$ is spacelike, yet non-singular,
generating a singularity-free black hole.  However, for such
a geometry, energy densities in the vicinity of the center
are necessarily exotic, since their constituents cannot be timelike. 
Alternatively, one can directly develop dynamic geometries for which
the center remains timelike perpetually, yet contain
transient bounded trapped regions.  
The exterior of such a geometry behaves similarly to a transient
black hole. 
It should be straightforward to examine unitarity and information
dynamics everywhere since the spacetime is horizonless and singularity-free.
Such a geometry, referred to henceforth as a
\emph{transient black object}, will be examined in what
follows.  

It should be noted that in certain circumstances, black objects
within which the center remains perpetually timelike nevertheless develop horizons,
thus also becoming black holes.  For instance, if the black object
does not evaporate away, a region develops within the object for which
outgoing lightlike trajectories reach $t=\infty$, but not
exterior lightlike future infinity, thus implying
a horizon.  Timelike trajectories interior to the trapped region
cannot escape, but exterior observers can fall in and become trapped.  However,
one cannot have a \emph{transient} black hole without developing
a spacelike center $r=0$.  Therefore, the transient black object considered in this paper is
distinct from a transient black hole in that the trapped information can
be recovered in its ``original" form in the future, in principle.

\subsection{Accretion \label{subsec:Accretion}}
\indent

One of the motivations for this model was the fulfillment of
energy conditions over the broadest possible region of
spacetime.  
Classical gravitating systems are expected to satisfy various energy
conditions everywhere.  However,
quantum systems exhibit spacelike coherent
behaviors, which can violate these conditions. 
Such violations are necessary for the evaporation of
black holes \cite{Birrell}.	

The null energy condition (NEC) and weak energy condition (WEC) state that the quantity
\be
\mathcal{I}_{null/weak}
 \equiv -u_{null/weak}^\mu \, T_{\mu \beta} \, u_{null/weak}^\beta  
\ee 
should be non-positive ($i.e.$ $\mathcal{I}_{null/weak}\le 0$), for lightlike (in the NEC case) and timelike (in the WEC case) observer
4-velocities, where $T_{\mu \beta}$ refers to the energy-momentum tensor. 
The dominant energy condition (DEC) refers to the 4-momentum $p_{source}^\mu$ of the matter distribution as seen by the observer with 4-velocity $\vec{u}_{obs}$,
given by
$p_{source}^\mu \equiv -{T^\mu}_\beta \, u_{obs}^\beta$. 
The DEC requires that this 4-momentum
satisfy 
\be 
\mathcal{I}_{observer}^{DEC} \equiv \vec{p}_{source} \cdot \vec{p}_{source} \le 0
\ee 
for all observers, where the dot product is defined by the metric of the geometry.  The DEC implies the WEC and NEC, so it provides the most significant constraint.  Matter distributions that do not satisfy the DEC are described herein as ``exotic."

The dominant energy condition for this geometry is satisfied if  
\be
0 \le  {\partial \over \partial ct} \, R_M (ct,r) \le  
2 \sqrt{R_M (ct,r) \over r} \, {\partial \over \partial r} \, R_M (ct,r) .
\ee
A solution to this non-linear relation satisfying the
energy conditions, $ R_M^{EC}$, is given by
\be
R_M^{EC} (ct,r)~=~ {4 \over 9} \, {r^3 \over (ct_B - ct) ^2} .
\label{RMEC}
\ee
When substituted into Einstein's equation, this form generates
a pressureless collapse of matter whose edge will be referred to as $r_{exterior}(ct)$.  
During accretion, the exterior surface satisfies the equation
\be
R_M(ct,r=r_{exterior}) ~=~ R_{So} ~\equiv~
{2 G_N M \over c^2},
\ee
where $M$ denotes the total mass in the cosmology. 
$R_{So}$ represents the Schwarzschild radius for a static
geometry with mass $M$. 
Thus, during the period of accretion only,
\be 
R_M(ct,r) ~=~ \left \{
\begin{array}{ll}
R_M^{EC} (ct,r) & \textnormal{interior, }  R_M^{EC} (ct,r)<R_{So} \\
R_{So}    & \textnormal{exterior, static vacuum.}
\end{array} \right .
\ee
The exterior surface $r_{exterior}(ct)$
can be shown to collapse at a sub-luminal rate. 
The accretion will be assumed to continue until the
exterior surface reaches a bounce scale
$L_{bounce}$ defined by micro-physics
that will be described in the next subsection.

\subsection{Evaporation \label{subsec:Evaporation}}
\indent

A region in spacetime within which any outgoing lightlike trajectory will have
decreasing radial coordinate initiates once $R_M (ct_{dark}, r=R_{So})=R_{So}$, $i.e.$ the matter has collapsed within its Schwarzschild radius. 
This defines the time of the onset of a trapped region as $t_{dark}$.  At
that time, evaporation due to quantum effects is assumed to begin.     
However, the collapse will continue indefinitely unless some microscopic effect
prevents complete degeneracy or the formation of a singularity. 
It will be assumed that such micro-physics contains a fundamental length scale
limiting gravitational degeneracy. 
This finite, arbitrarily small, scale will be chosen here to have a fixed
value $L_{bounce}$, beyond which microscopic pressures and quantum
non-locality prevent further collapse. 
The functional behavior of $R_M$ interior
to $L_{bounce}$ must then be of a form that prevents the formation
of a singularity, $i.e.$ such that $Lim_{r \rightarrow 0}{R_M(ct,r) \over r} < \infty$.  If this limit attains
a value greater than unity, a horizon will form, creating a non-singular \emph{black hole}.  Otherwise,
the center $r=0$ will remain timelike everywhere, and if a trapped region forms, the
geometry will manifest a generic \emph{black object}.  In the exterior, both types of dark
geometries are quite similar prior to complete evaporation. 
Since there is no spacelike center, a sturdy system on a timelike trajectory can in principle be detected after evaporation of the black object.

The dynamics of information and the global causal structure of a dynamic
black object is the subject of this investigation. 
During black object evaporation,
the exterior region is delineated by the outermost
trapping surface $R_S (ct)$, which is defined in terms of the
interior mass of the the black object that has yet to evaporate
\be
R_S (ct)  \equiv {2 G_N M(ct) \over c^2}.
\ee
The dynamics describing the evaporation of this radial surface scale
will be motivated using Hawking-like thermal radiation rates expected
from a quasi-static geometry.  The emission of such radiation from trapping surfaces has been discussed in the literature \cite{Visser01}. 
The rate of interior mass change
is expected to be of the form
\begin{displaymath}
\dot{M} c^2 = {\textnormal{characteristic energy} \over \textnormal{emitted quantum}} \times
{\textnormal{number of quanta emitted} \over \textnormal{unit time}},
\end{displaymath}
where the dot indicates a derivative with respect to $ct$.  
The energy of the emitted quantum is expected to be defined by the
spatial extent of the radial surface scale $R_S (ct)$, which likewise
defines the temperature of a quasi-static geometry.  For generality,
an arbitrarily small additional microscopic scale $\delta_R$ will be included to prevent
an indefinitely large energy, giving a form for this term of the
order ${\hbar c \over R_S (ct) + \delta_R}$. 
The rate of emission of the quanta for fiducial observers in the static geometry
is expected to be of the order of one quantum per unit
Rindler time \cite{LSJLBlackHoles}, which is likewise inversely proportional
to the radial surface scale.  Again, for generality, an arbitrarily small
microscopic scale $\delta_T$ will be added, resulting in a rate of
emission ${\kappa \over R_S (ct)+\delta_T}$. where $\kappa$ is a dimensionless number of order one.  Thus, multiplying
by $2 G_N \over c^2$, the dynamics of the surface scale will be taken
to satisfy
\be
\dot{R}_S (ct) ~=~- {2 L_{Planck} ^2 \over R_S (ct) +\delta_R} \, 
{\kappa \over R_S (ct)+\delta_T}.
\label{HawkingRate}
\ee

After the interior mass has decreased to a point where the radial surface scale is equal
to the bounce scale $L_{bounce}$, the geometry no longer
contains a trapped region, and the black object vanishes. 
Subsequently, all outgoing lightlike trajectories will have
increasing radial coordinates.
If there is no further quantum decay, a remnant of this mass
will remain.  However, for the present
investigation, the core region will be assumed to continue
quantum decay through massless quanta.  
The gravitationally
stabilized quantum decay will be presumed to generate
quanta that satisfy quantum measurement constraints
as well as geometric consistency, as will be discussed shortly. 
During this decay, the exterior surface scale delineating the
interior and exterior regions will take the fixed value
$L_{bounce}$.  \footnote{Alternatively, if this scale were to maintain a time-dependent fractional value relative to $R_S (ct)$, the black object would continue thermal decay until all interior mass has evaporated. The development of such a model is quite straightforward, and its features do not exhibit significant modification from the model herein explored.} 
The radial surface scale and the the exterior surface
scale are demonstrated in Figure \ref{RsPlot}.
\twofigures{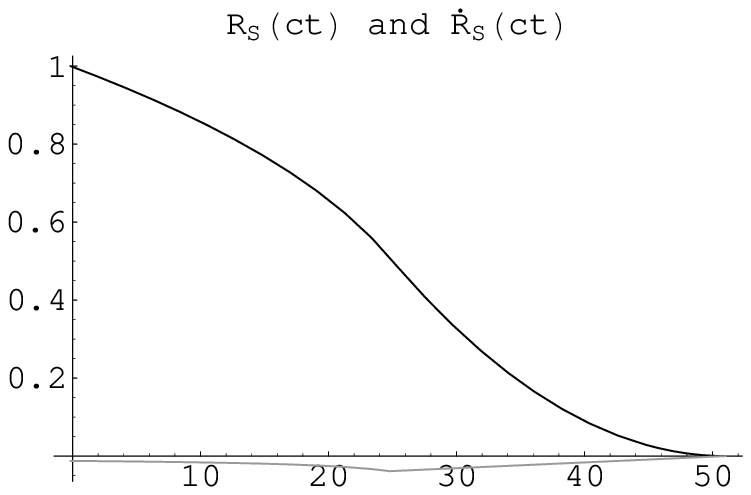}{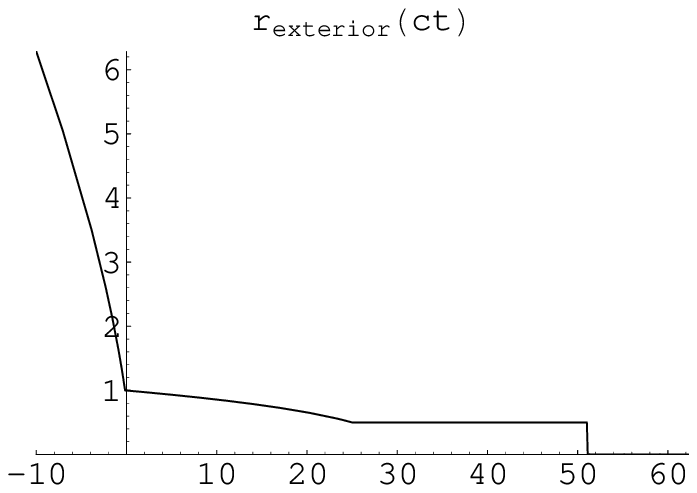}{Plots of $R_S(ct)$
and $\dot{R}_S (ct)$ (left), and $r_{exterior}(ct)$ (right). 
The black curve on the left plot represents the radial surface
scale, while the negative gray curve represents its derivative.}
The particular form chosen for the decay is given by
$R_S (ct) = A \, (ct-ct_{final})^2$, where $ct\le ct_{final}$ and the constant $A$
is chosen to smoothly match the final rate of thermal evaporation
with the initial rate of decay.  The quadratic form allows a smooth transition to Minkowski space.

In the exterior region, changes in the interior mass will be presumed to
be communicated via collections of massless quanta that carry energy and change the geometry.
The quanta collectively carry sufficient energy to change the interior
radial surface scale by $\delta R_S$, where $\delta R_S<0$. 
The quanta then propagate through
a static affine space with lesser interior radial surface scale $R_S + \delta R_S$.
If this is done consistently, all energy conditions will
be satisfied in the exterior. 

Quantum measurement constraints can be estimated by examining
the energy-at-infinity $\delta E$ 
carried per emission by the outgoing quanta.  The fact that $\delta R_S = \dot{R}_S \, \delta ct$,
implies that $\delta E = -{c^4 \over 2 G_N}\delta R_S = {\hbar c \over R_S (ct) +\delta_R}$.
Incorporating the uncertainty principle
$\delta E ~ \delta t \ge {\hbar \over 2}$ implies that the rate of decrease in the radial surface
scale (for $\delta_R \simeq \delta_T$) has a lower limit given by
$\delta R_S ~ \delta ct_{emissions} > L_{Planck}^2$. 
However, there is also an upper limit upon the rate of evaporation/decay, due to
geometric consistency and causality. 
Specifically the rate of thermal evaporation must be dynamically consistent so that energy carrying quanta do not
leave spacetime flat enough that subsequent quanta eventually catch up. 
Indeed, a rapid enough evaporation would violate geometric consistency. 
The chosen decay form generates successive outgoing massless quanta
that asymptotically have separations demonstrated in Figure \ref{DrSpread}. 
\onefigure{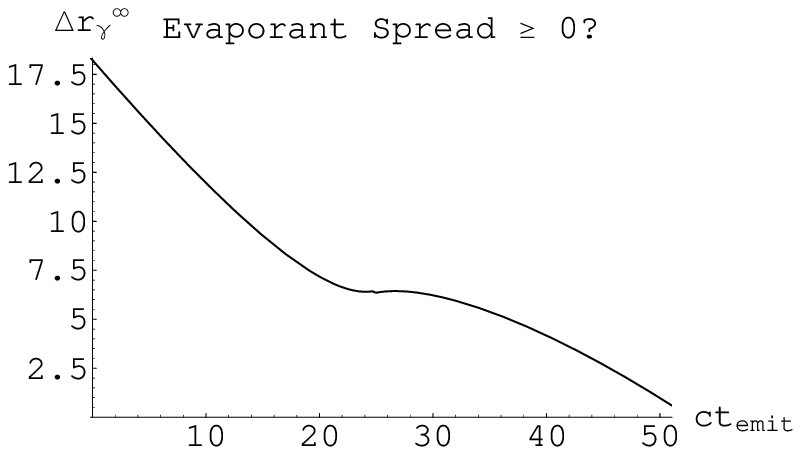}{Asymptotic spread of sequentially emitted energy-carrying quanta.}  
In the figure, geometry-changing quanta are emitted from the emission surface at
small fixed intervals and propagated as outgoing quanta consistent with
the metric (\ref{BOmetric}) on null trajectories
satisfying $\dot{r}_Q = 1-\sqrt{R_M / r_Q}$.

Therefore, the specific model developed incorporates the following points:
\begin{itemize}
\item The geometry-changing quanta must emit exterior to the background radial
surface scale $R_S^{bg}$ through which they propagate, if they are to
transport mass away;
\item The geometry-changing quanta will thus propagate through a background geometry
with surface scale $R_S + \delta R_S$ carrying energy $\delta E= -\frac{c^4}{2G_N}\delta R_S$;
\item The emissions communicate the transport of energy as a lightlike
outgoing energy that includes gravitational binding;
\item Since the exterior radial mass scale incorporates the geometry-changing
massless quanta, all energy conditions are expected to
be satisfied in the exterior.
\end{itemize}
Each massless geometry-changing quantum ``sees" a static geometry; it merely propagates through the geometry left behind by its immediate predecessor. 
Exterior outgoing quanta emitted at $(ct_o,r_o)$,
located at $r_\gamma (ct)$, and propagating through a background geometry
parameterized by $R_S^{bg}$ satisfy
\be 
ct - ct_o = r_\gamma (ct) - r_o + 2 \sqrt{R_S^{bg}} (\sqrt{ r_\gamma (ct)}-
\sqrt{r_o}) + 2 R_S^{bg} \, log \left (
{\sqrt{ r_\gamma (ct)}-\sqrt{R_S^{bg}} \over \sqrt{r_o}-\sqrt{R_S^{bg}}}
\right ) .
\ee
The geometry-changing quanta generated by evaporation or decay will thus
propagate through a local background geometry characterized by
by $R_S^{bg}(ct_o)=R_S(ct_o) + \delta R_S(ct_o)$.  In order to calculate
the radial mass scale at an arbitrary exterior point $(ct,r)$, the
retarded event of emission $(ct_o,r_o)\equiv (ct_{ret},r_{o}(ct_{ret}))$
must be determined. 
During evaporation, the exterior emission scale will be chosen to
satisfy $r_{o}(ct_{ret})=R_S(ct_{ret})+\delta_{stretch}$ (where $\delta_{stretch}$ can be arbitrarily small because the quanta propagate on a background of scale $R_S + \delta R_S$ as opposed to $R_S$ itself).  

The radial mass scale at a general exterior point will therefore be
given by
\be
R_M (ct,r) ~ \equiv~ R_S (ct_{ret}(ct,r)) \quad , \quad \textnormal{$r>r_{exterior}(ct)=R_S(ct)$}.  
\ee
This form then ensures a causal propagation of the evaporation of mass
from the core region.  During evaporation the core region is chosen to maintain its previous $r$-dependence from the accretion period, such that  
\be
R_M (ct,r) ~ \equiv~ {R_S (ct) \over R_S (ct_{dark})} 
R_M^{EC} (ct_{dark},r) \quad , \quad r\le L_{bounce}
\ee
where the pressureless form satisfying energy condtions $R_M^{EC}$
is given by Eqn. (\ref{RMEC}).  The region between the exterior and the
bounce scale $L_{bounce}$ will be assumed to maintain the spatially coherent
form $R_S (ct)$.  The chosen forms smoothly match the behaviors during the transition from
accretion to evaporation.

As previously mentioned, this form for the radial mass scale has been chosen
so as to satisfy energy conditions in the broadest region of the spacetime. 
Plots of the evolution of the radial mass scale, along with local measures
of fulfillment of the energy conditions, are represented in Figure \ref{BOEC1}.
\fourfigures{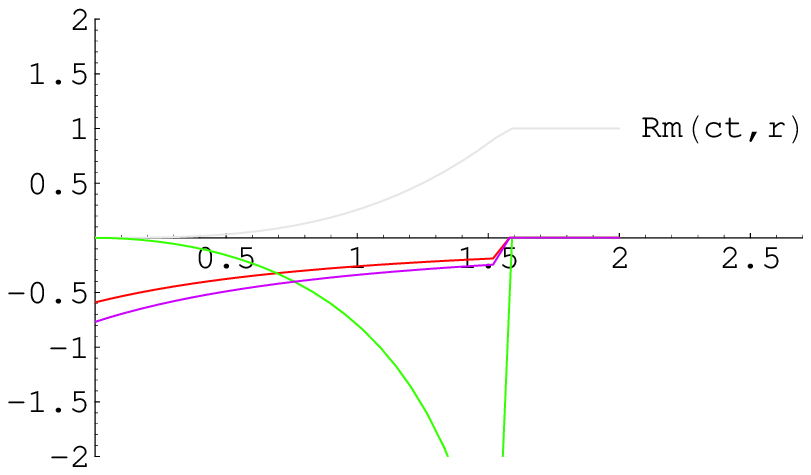}{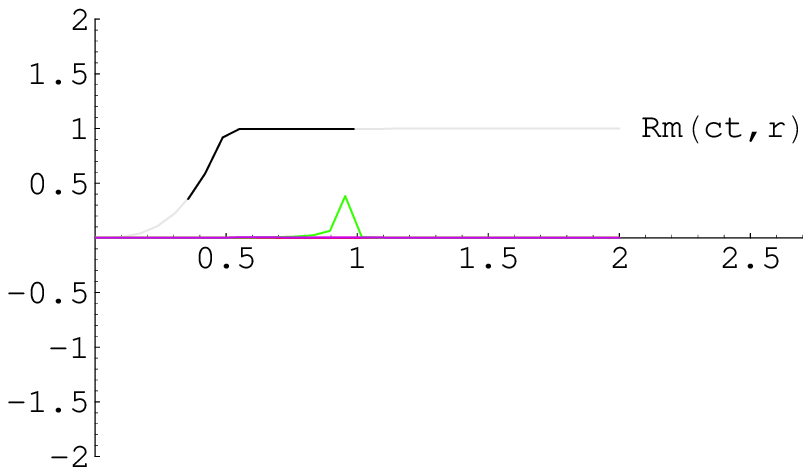}{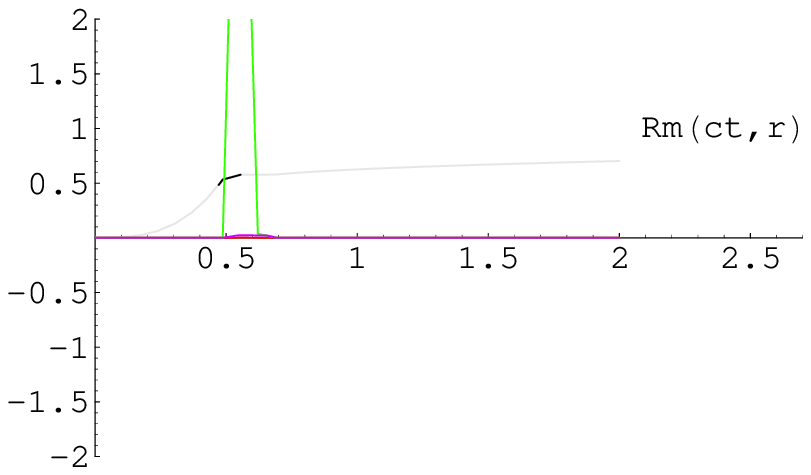}{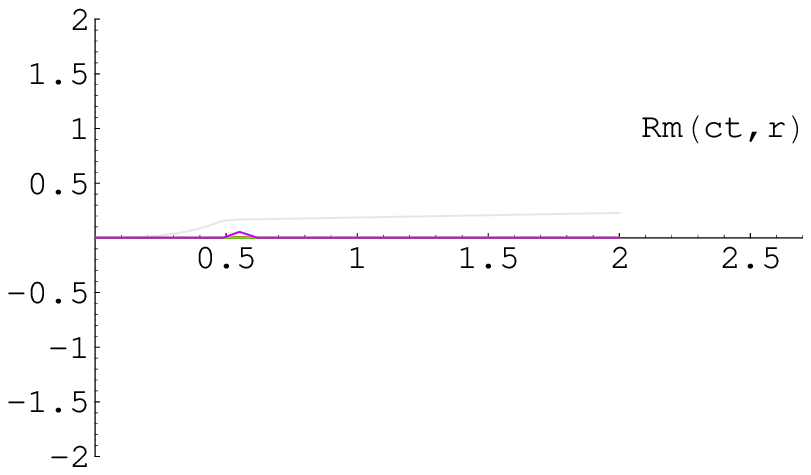}{Radial mass scale and
energy conditions for late accretion (upper left), final accretion with evaporation (upper right),
mid-evaporation (lower left), and late decay (lower right).  Both axes are measured in units of $R_{So}$. 
The radial mass scale is shown in gray with a black overlay of trapped region. 
The DEC invariant ${I}_{observer}^{DEC}$ is shown for radial observer motions in red and for azimuthal observer motions in green. The NEC invariant $\mathcal{I}_{null}$ is shown in purple.  The energy conditions are satisfied for ${I}_{observer}^{DEC}\le 0$ and $\mathcal{I}_{null}\le 0$, respectively.}  
The diagrams in the figure depict snapshots of these quantities during late accretion,
the final stage of accretion after the black object has formed and evaporation has begun,
the late stage of thermal evaporation, and the late stage of a gravitationally suppressed
quantum decay of the remnant.   

The gray curves in Figure \ref{BOEC1} represent the spatial
dependence of the radial mass scale at the given time.  If there is a trapped region,
it is represented by a black segment upon the radial mass scale curve.  It is clear from the diagram that the trapped region has both an inner boundary $R_{TS}^-$ and an outer boundary $R_{TS}^+$.  
Curves are also demonstrated depicting local values for the DEC invariant ${I}_{observer}^{DEC}$ for radial observer motions (red), ${I}_{observer}^{DEC}$ for azimuthal observer motions (green), and the NEC invariant $\mathcal{I}_{null}$ (purple), using a convention in which the DEC and NEC are satisfied when ${I}_{observer}^{DEC}\le 0$ and $\mathcal{I}_{null}\le 0$, respectively.  
All energy conditions are everywhere satisfied
prior to the beginning of evaporation, as expected.  Also, energy conditions
are satisfied everywhere in the exterior region $r>r_{o}(ct)$, implying that any measurements in the
exterior will never detect exotic energy forms.  The snapshots demonstrate
violation of the energy conditions only within the region of spatial coherence
where production of the quanta occurs.  Such violations of energy conditions
should be expected within the trapped regions, since these are also classically
forbidden regions.

\subsection{Construction of lightlike curves \label{subsec:LightCurves}}
\indent

The exploration of the dynamic
features of a system is straightforward on a conformal
diagram.  Unfortunately, many useful dynamic geometries
such as this transient black object do not
afford a direct calculation of a set of conformal
coordinates.  A general technique is therefore necessary to construct Penrose diagrams
in complicated geometries.

The technique
relies only upon constructing lightlike surfaces for the
given metric, in this case given in Eqn. (\ref{BOmetric}). 
Once those null geodesics have been generated, the
conformal coordinates can be labeled $(v,u)$,
based upon the correspondence of the lightlike curves on reference
hypersurfaces (in this case, the exterior surface and ultimately $skri^\pm$). 
Conformal spacetime coordinates $(ct_*,r_*)$ can be introduced such that  $v=ct_* + r_*$ and
$u=ct_* - r_*$.  For ingoing null geodesics,
the required equation labeled by conformal
coordinate $v$ takes the form
\be
\dot{r}_v ~=~ -1
 - \sqrt{R_M \over r_v} \, ,
\label{rvEqn}
\ee
while outgoing null geodesics labeled by $u$ satisfy
\be
\dot{r}_u ~=~ 1
 - \sqrt{R_M \over r_u} \, .
\label{ruEqn}
\ee
For this geometry, ingoing lightlike trajectories labeled by
$v$ initiating on past lightlike infinity have access to all regions
of spacetime.  Outgoing lightlike trajectories labeled by
$u$ terminating on future lightlike infinity likewise have access to all regions
of spacetime, since there is no horizon.

Ingoing lightlike trajectories defining the
conformal coordinate $v$ for the transient black
object are demonstrated in Figure \ref{BOPhotonIn}.
\onefigure{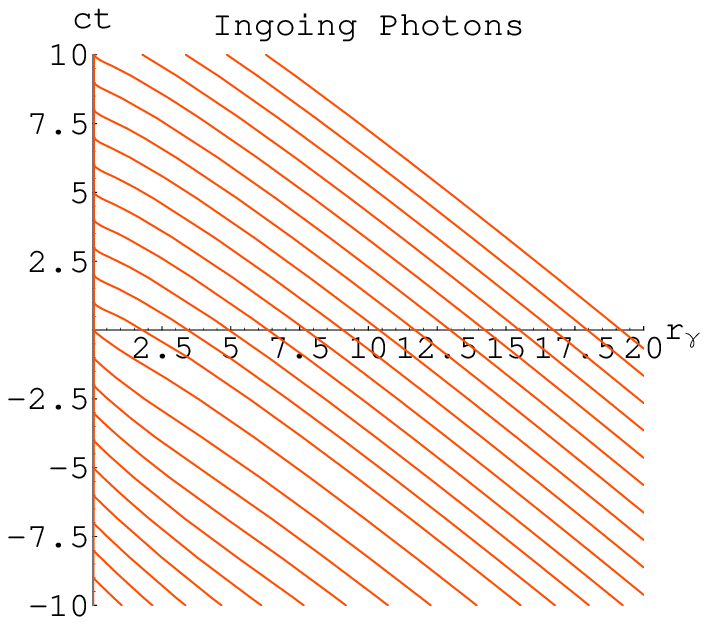}{Ingoing lightlike trajectories.
Results are displayed in units of $R_{So}$.}
The ingoing photons are chosen to terminate at $r=0$, temporally
separated in units of the Schwarzschild radius of the geometry $R_{So}$.
For the chosen parameters, the trapped region first develops
at $ct \simeq -0.167 R_{So}$, and vanishes at 
$ct \simeq 25 R_{So}$.  Accretion ends at 
$ct \simeq +0.167 R_{So}$,
and final decay occurs at $ct \simeq 51.05 R_{So}$.
Ingoing photon trajectories are seen to propagate through
essentially flat spacetime until they approach the trapped
region, within which their radial coordinates decrease
more rapidly than in Minkowski spacetime. 
In the static exterior of the past, ingoing massless
quanta satisfy
\be  
ct_o - ct = r_\gamma (ct) - r_o - 2 \sqrt{R_{So}} (\sqrt{ r_\gamma (ct)}-
\sqrt{r_o}) + 2 R_{So} \, log \left (
{\sqrt{ r_\gamma (ct)}+\sqrt{R_{So}} \over \sqrt{r_o}+\sqrt{R_{So}}}
\right ) .
\ee
Thus, the outgoing lightlike surface of constant $u$ communicating the beginning of evaporation
can serve as the exterior surface of correspondence for assigning
the parameter $v$ during evaporation. 

Outgoing lightlike trajectories defining the
conformal coordinate $u$ for the transient black
object are demonstrated in Figure \ref{BOPhotonOut}.
\onefigure{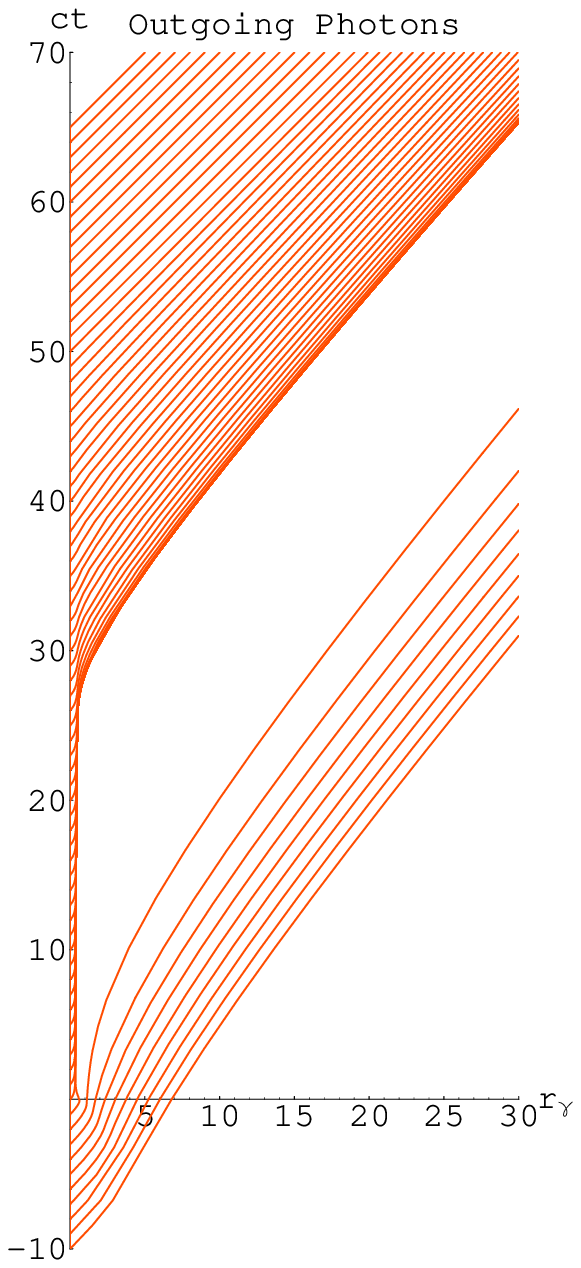}{Outgoing lightlike trajectories (that originate at $r=0$). 
Results are displayed in units of $R_{So}$.}
The outgoing photons are emitted from $r=0$, temporally
separated in units of the radial mass of the geometry $R_{So}$.
Photons emitted after the final decay of the black object propagate
through flat spacetime.
Likewise, photons emitted in the distant
past propagate through nearly Minkowski spacetime.
However, photons emitted just prior to the formation of the black object initially
propagate with increasing radial coordinate, then have decreasing
radial coordinate as the
trapped region forms.  Photons emitted during the lifetime of the black
object (while there is a trapped region) also initially propagate with increasing
radial coordinate, but slow as they approach the inner
surface delineating the trapped region, being temporarily
trapped within the innermost core region.  It should be noted that
\emph{all} outgoing photons eventually will reach lightlike future
infinity in this transient black geometry.

\section{Conformal diagram of the transient black object
\label{sec:BOPenrose}}
\indent

Light-cone conformal coordinates $(v=ct_*+r_*,u=ct_*-r_*)$ parameterized
in the previous section will be made compact using the
identification 
\be
\begin{array}{l}
Y_\rightarrow =[tanh({ct_*+r_* \over scale})-
tanh({ct_*-r_* \over scale})]/2 \, , \\ \\
Y_\uparrow =[tanh({ct_*+r_* \over scale})+
tanh({ct_*-r_* \over scale})]/2 \, .
\end{array}
\label{PenroseCoords}
\ee
The coordinates $(Y_\rightarrow, Y_\uparrow)$ can be used
to construct the Penrose diagram
of this geometry with a transient black object. 
The geometry indeed has
some finite period with real solutions of Eqn. (\ref{TrappingSurfaceEqn})
for the outer and inner surfaces of the
trapped regions 
satisfying $R_{TS}^- \not= R_{TS}^+$,
with the chosen parameters for the metric (\ref{BOmetric}).

The conformal diagram, with significant transitional
epochs indicated, is shown in
Figure \ref{BOPnrFeat}. 
\twofigures{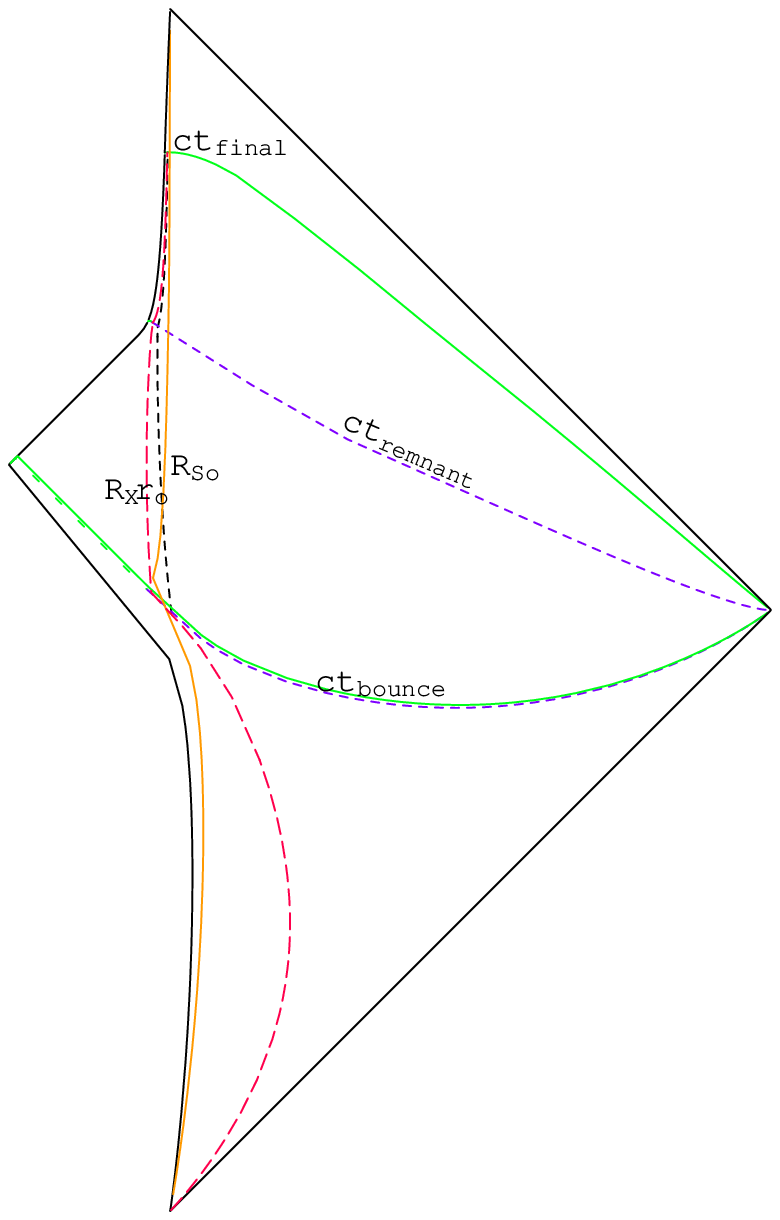}{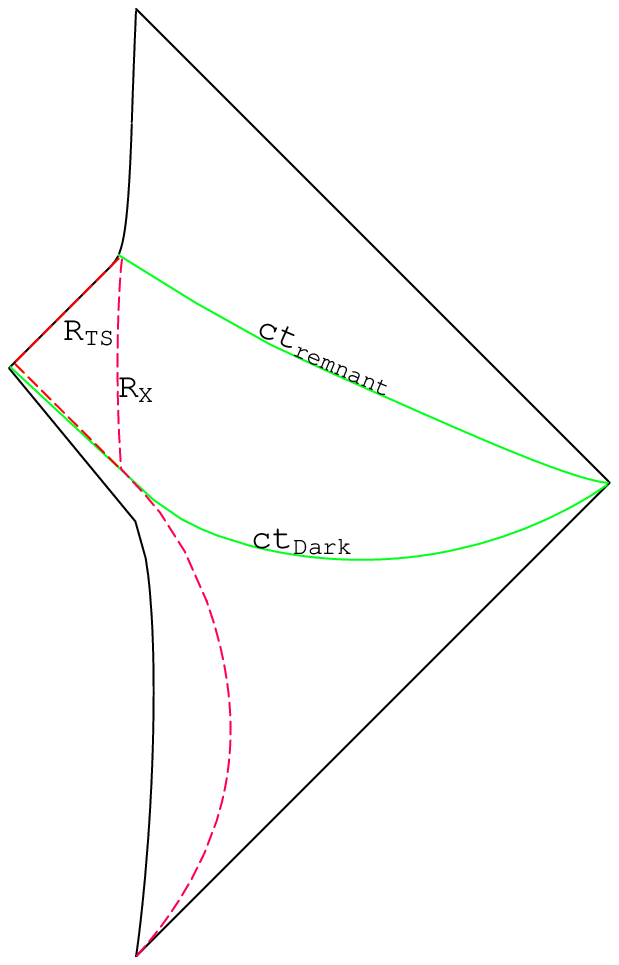}{Conformal diagrams of the transient
black object, demonstrating dynamic features of interest.  The edge $r_{exterior}$ has been abbreviated $R_X$.  The diagram on the right 
emphasizes the boundaries of the trapped region.  Once the black object forms, the outer boundary $R_{X}$ is always timelike, while the inner boundary $R_{TS}$ is spacelike during evaporation.}
The diagram is bounded from the left by the timelike
center $r=0$, from the lower right by past lightlike
infinity $skri^-$, and from the upper right by future
lightlike infinity $skri^+$.  There are no horizons on the diagram. 
The static radial mass (Schwarzschild radius) of the geometry depicted by
the (orange) curve labeled $R_{So}$ sets the scale of the diagram. 
The center of the conformal coordinates
$(Y_\rightarrow =0, Y_\uparrow =0)$ is chosen to
correspond with the coordinate $(ct=0, r=R_{So})$. 
The timelike surface that delineates the exterior region of the
geometry is depicted by the (red) dashed curve labeled
$R_X$.  During accretion, this curve represents the radial
coordinate within which all mass in the spacetime is
interior.  After evaporation begins $(ct>ct_{dark} \approx
ct_{bounce}$), $R_X$ represents the
radial scale within which mass that has yet to evaporate
or decay away is contained.  The spacelike surface showing
the beginning of evaporation ($ct_{dark}$) is represented by the
dashed (green) curve just prior to the solid (green) curve labeled $ct_{bounce}$ in the diagram on the left.  The latter depicts the end of accretion. 
The geometry-changing quanta emit from the surface labeled $r_o$,
depicted by the black dashed curve.  The dashed spacelike surface labeled
$ct_{remnant}$ depicts the end of thermal evaporation, and the
beginning of decay of the remnant.  Finally, the solid
(green) spacelike curve labeled $ct_{final}$ depicts the end
of decay of the remnant of evaporation.  The region of the spacetime
above the communication of the end of decay is flat.  The diagram on the right demonstrates the boundaries of the trapped region.

Fixed-coordinate surfaces in the geometry are demonstrated in
Figure \ref{BOPenrCoord}. 
\twofigures{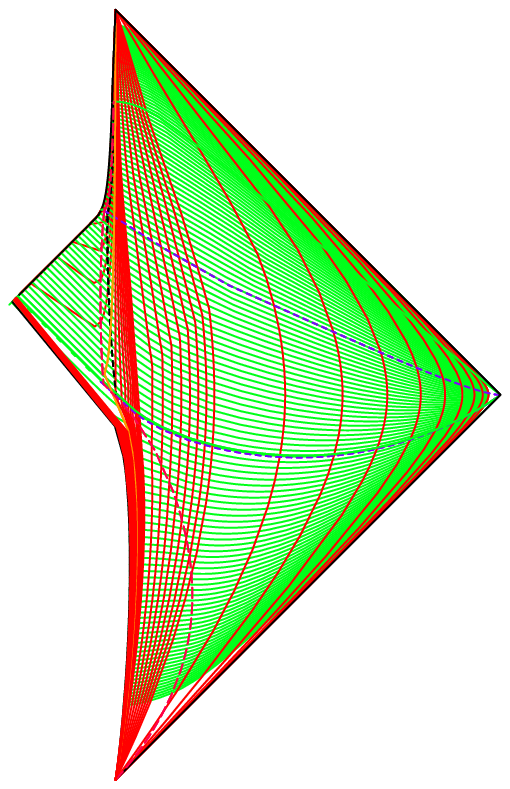}{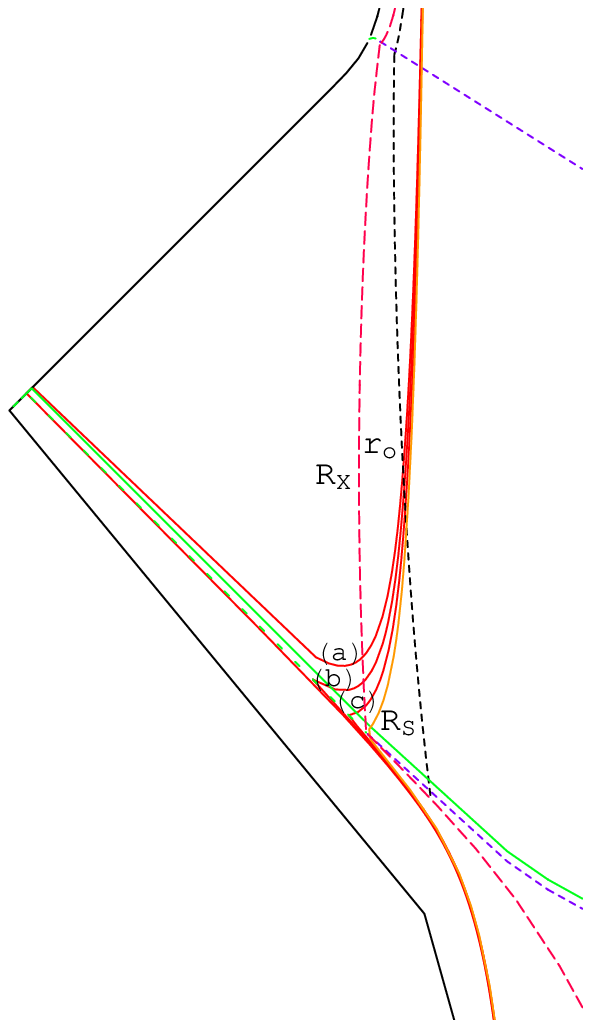}{Fixed-$ct$ and fixed-$r$ surfaces on a conformal diagram
for the transient black object.  Left: Spacelike surfaces
of fixed $ct$ are represented by green curves graded in units
of $R_{So}$.  Curves of fixed radial coordinate $r$ are represented
by red curves graded in units of $0.1 R_{So}$ from the center out to
$2 R_{So}$, then integral values of this unit, then decades of this unit. 
Right: Close-up view of (a) $r=0.97 R_{So}$,
(b) $r=0.98 R_{So}$, and (c) $r=0.99 R_{So}$.}
One of the most useful characteristics of this type of metric is
that fixed-time coordinate surfaces are everywhere spacelike, and
are shared by a set of geometrically stationary co-movers. 
This structure of the geometry is apparent in the figure. 
However, fixed-radius surfaces are also spacelike
within the trapped region.  In the diagram on the left, fixed-radius
coordinate surfaces are everywhere timelike for $r>R_{So}$.  Also,
fixed-radius surfaces interior to the minimum value
of the inner boundary of the trapped region $r<R_{TS}^-$
are always timelike, including the center $r=0$.  The boundaries
of the trapped region represent surfaces for which outgoing
lightlike trajectories are momentarily stationary in the radial coordinate. 
Those fixed-radius surfaces that lie within the trapped region
demonstrate transition from timelike to spacelike, then back
to timelike behaviors exhibited in the leftmost region of the diagram. 
For clarity, this region is expanded in the diagram on the right of
Figure \ref{BOPenrCoord}.  In this diagram, fixed-radius surfaces
near the static mass scale $R_{So}$ are depicted.  The slope of each
of the surfaces $r=constant$ is timelike prior to the formation of the
black object, becomes unity as the innermost surface of the trapped
region $R_{TS}^-$ crosses that coordinate prior to the fixed-coordinate surface becoming
spacelike, again becomes unity as the outermost surface
of the trapped region $R_{TS}^+ = R_S (ct)$ crosses that
coordinate, and is ultimately again timelike.  This behavior
has been observed in prior explorations of coordinate surfaces in transient
trapped geometries \cite{JLPS2010, JLTF2010}.

The behaviors of the outgoing null trajectories
near the black object displayed in Figure \ref{BOPhotonOut}
demonstrate that the outgoing communications from
\emph{any} system just before crossing the trapping surface are
temporarily held near that surface. 
This is true for infalling systems at any time throughout
the existence of the dynamic black object.  However, all of these
communications will eventually reach a nearby exterior observer
after a finite time.  In contrast, the analogous outgoing lightlike
trajectories near a Schwarzschild horizon
will release communications that only
reach the exterior observer after an
indefinitely long period of time. 
The next section will examine the dynamics of the
release of these communications as systems
enter the black object.

\setcounter{equation}{0}
\section{Dynamic information from
trapped region-traversing observers
\label{DynamicInfo}}
\indent

In order to examine the information dynamics of
the transient black object, a standard exterior observer will
be chosen to be geometrically stationary, thus sharing
proper time with that of the asymptotic observer ($t$)
parameterized in the metric. 
The initial conditions of this nearby observer can be chosen
such that the observer avoids ever encountering the trapped region. 
This co-moving observer, whose trajectory
is given by the outermost bold trajectory in Figure \ref{SpTmPhotons}
will serve as the observational platform examining gravitating
systems entering the trapped region.  The trajectory
was chosen to correspond
to that of a stationary observer at $r=R_{So}$ after the remnant has
completed its decay.

\twofigures{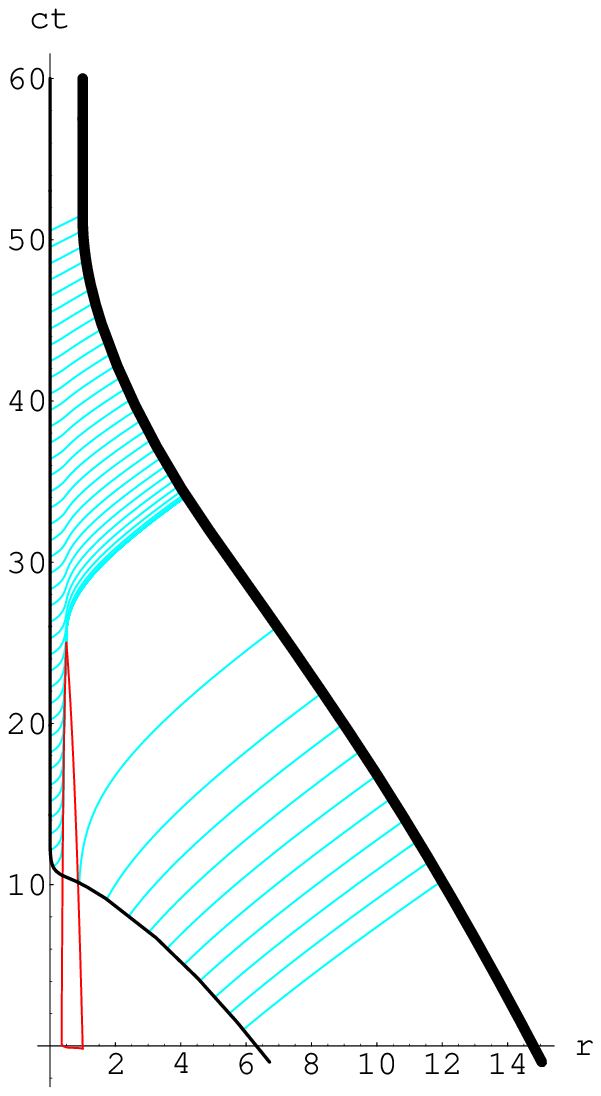}{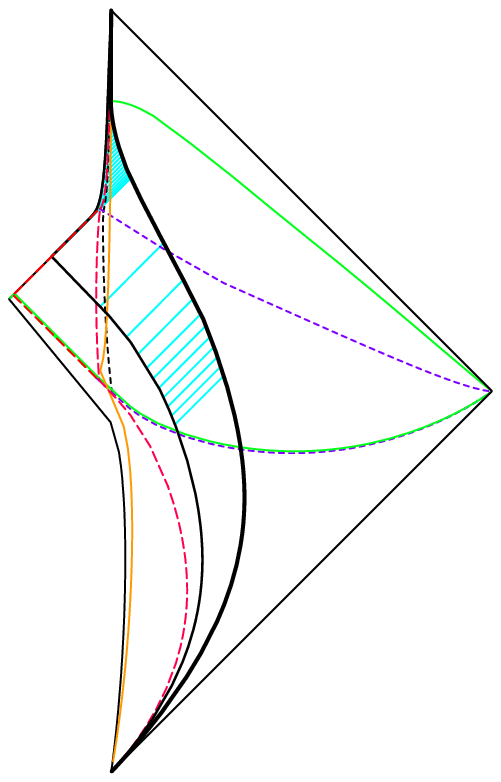}{Standard photons that are emitted by
a geometrically stationary infalling source
and detected by a nearby external observer.  A standard spacetime
diagram (which displays the boundaries of the trapped region) is represented on the left, and the conformal diagram is represented on the right.}

The observed infalling system will emit ``standard" frequency photons
at a ``standard" rate in its proper coordinate frame
of reference.  In what follows, the emitter will be a
freely falling, geometrically stationary system that
reaches $r=R_{So}$ at time $ct=10 R_{So}$ (which is during evaporation), then
falls through the trapped
region.  This infalling emitter
is depicted by the innermost bold trajectory in Figure \ref{SpTmPhotons}. 
The emitter approaches, but never reaches, the center $r=0$.
In the figure, the outgoing photons are emitted at $ct=R_{So}$, $ct=2R_{So}$,
$ct=3R_{So}$, $etc$. 
There is an extended period of time during which only highly redshifted
photons are observed from the emitter.  This especially occurs between the communication of
the emitter crossing into the trapped region and the communication of the end of the black object. 
After the emitter passes through the inner boundary of the trapped region, all outgoing communications are temporarily trapped inside that boundary, prevented from entering the trapped region.  These photons ``bunch up" within the inner surface of the trapped region, but are rapidly released after the trapped region vanishes, in a manner similar to that illustrated in Figure \ref{BOPhotonOut}.  Eventually, all emitted
communications will be received by the observer.

\subsection{Frequency shifts of quanta emitted by infalling emitters.}
\indent

In order to examine the redshift of the emitted quanta, the null trajectories (\ref{ruEqn}) that are obtained directly from the metric are not sufficient; the null geodesic equation
\be
{du^\beta \over d \lambda} + 
\Gamma_{\mu \nu}^\beta u^\mu u^{\nu} =0 ,
\ee
is required (here $\lambda$ is the affine parameter). 
The geodesic equation is used to calculate the 4-velocities for massless quanta.  Subsequently, it is seen that the quantity $-\vec{u}_{obs} \cdot \vec{u}_\gamma $,
the observed value of the temporal component of the 4-velocity
 of a radially outgoing photon, satisfies
\be
-\vec{u}_{obs} \cdot \vec{u}_\gamma =
\left (  u_{obs}^{ct} \mp \sqrt{ (u_{obs}^{ct})^2 -1}    \right ) \, 
 u_\gamma^{ct},
\label{ObservedPhotonu0}    
\ee
where the $\mp$ sign refers to outgoing/ingoing observers.  
This is directly proportional to the observed energy of
the photon.  For a geometrically stationary observer, $u_{obs}^{ct}=1$, and
the observed photon temporal component is 
seen to be simply $u_\gamma^{ct}$.
The relationship (\ref{ObservedPhotonu0}) 
can likewise be used to express $u_\gamma^{ct}$
in terms of its value $(u_\gamma^{ct})_{proper} = -\vec{u}_{*} \cdot \vec{u}_\gamma$ in the proper frame * of the emitter:  
\be
u_\gamma^{ct}={(u_\gamma^{ct})_{proper} \over 
 u_*^{ct} \mp \sqrt{ (u_*^{ct})^2 -1}  } ,
\ee 
where the $\mp$ sign refers to outgoing/ingoing emitters.
Since both the observer and the emitter are geometrically stationary, 
the geodesic equation directly calculates the
observed frequency changes of the gravitating photons.

The rate of reception of information emitted from the infalling
emitter can be obtained by examining the component
$u_\gamma^{ct}$ for the outgoing standard photons. 
The standard frequency (in the frame of the source) for photons emitted will be chosen
such that  ${u_{\gamma} ^{ct}}_{standard}$=1.
The temporal behavior of $u_{\gamma} ^{ct}$ for standard
photons from the infalling emitter \emph{as measured by the
geometrically stationary
observer} is shown in Figure \ref{Redshiftu0}.
\twofigures{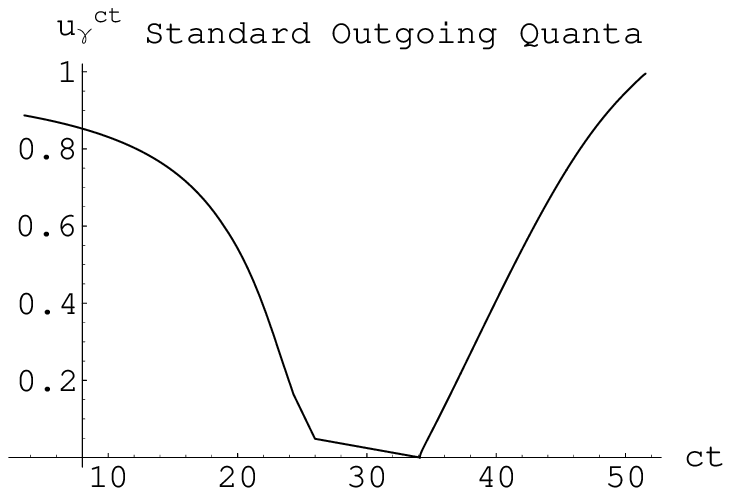}{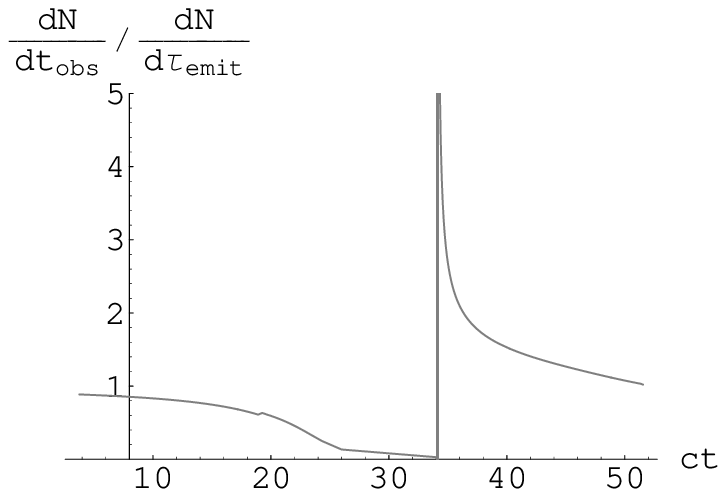}{Left: Observed $u_\gamma ^{ct}$ of 
standard photons as a function of observation time.
Right:  Ratio of rate of observation of detected quanta to the rate of standard emission.}
The figure demonstrates both the evolution of the frequency
of individual photons on the left, as well as the rate of observation
of photon emissions on the right. 
The rate of emission of photons is related to the interval
$d\tau_{emit}$ between successive photons.  Likewise, the
interval between the observation of those successive photons
by the co-moving observer $dt_{obs}$ relates to the rate
of detection.  Thus, prior to the disappearance of the trapped region, the ratio of the rate of
observation to the standard rate of emission
should equal the ratio of the observed frequency to the standard frequency.    
As a check for numerical accuracy,
this independent measure of redshift on the right of Figure \ref{Redshiftu0} indeed
functionally coincides with the geodesic calculation displayed on the left
prior to the communication of the end of the black object. 
However, after communications
from the emitter emerge following the evaporation of the trapped region, the
individual photons depicted in the left diagram are initially quite redshifted and ultimately approach
unit ratio from below, while the rate ratio in the diagram on the right is quite enhanced and ultimately approaches
unit ratio from above.  This should not be surprising, since such behavior is
apparent in Figure \ref{SpTmPhotons}.

The observation of infalling emitters through the trapped region of a transient
black object is somewhat different from the observation of infalling emitters through
the horizon of a transient black hole \cite{JLTF2010}.  For emitters falling into
transient black holes, any photons emitted by infalling systems before the horizon is crossed will continue
to be redshifted until those systems are seen to completely vanish as the black hole
itself fully evaporates away.  This is true independent of the actual time that the
system falls through the horizon, $i.e.$, all infalling emitters are seen to simultaneously
vanish as they traverse the vanishing horizon of a transient
black hole.  However, for the transient black object,  
\emph{all} emissions rapidly revert toward standard frequency and rate of emission as
the trapped region evaporates.  For a static black hole, emitters are
\emph{never} observed to traverse the horizon (see, for instance, reference \cite{LSJLBlackHoles},
page 23).

To complete this examination, a measure of the
power of emissions from the infalling system as observed
by the external observer will be developed.  The power should be  related
to the energy of each photon times the rate at which photons are observed. 
This power parameter measured for radially outgoing photons at
the radial coordinate of the external
observer is displayed in Figure \ref{Power}.
\onefigure{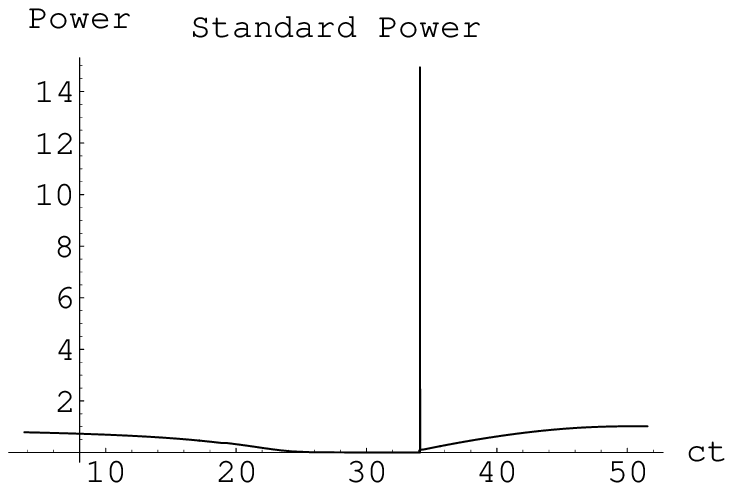}{Normalized power $P_{standard}$ received by the geometrically stationary exterior observer.}
After final decay of the remnant, and for very early times, this standard power takes the value of
unity.  Just as the trapped region vanishes, there is a considerable spike observed in this
measure of power.  

If the detector used by a single external observer is small enough
(or if there were an isotropic set of infalling emitters all radiating radially) the emissions are uniform over the observed solid angle. 
The actual detections are then expected to fall off with the inverse square of the
radial distance traveled by the photon, which for a geometrically
stationary emitter and observer is given by
$r_{obs} (ct_{observed})-r_* (ct_{emit})$.  A diagram of this
observed measure of ``intensity" is demonstrated in Figure \ref{Intensity}.
\onefigure{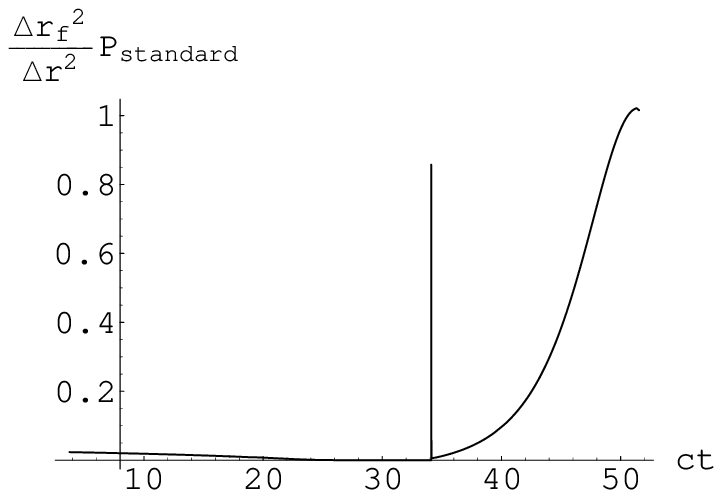}{``Intensity" measured by the exterior observer using a small detector.}
Early measurements of this intensity are smaller than later measurements because
of the greater distance between the emitter and the source.  This intensity is normalized
to take the value unity after the remnant has completely decayed.  At this time, both the observer and system are motionless since geometrically stationary objects have fixed spatial coordinates in Minkowski spacetime.  The figure demonstrates a spike in intensity as the trapped photons are released after the termination of thermal evaporation, which corresponds to the end of the black object.

\subsection{Entanglement of massless quanta.}
\indent

The absence of a spacelike center for the singularity-free black object allows
a direct exploration of the trajectories of gravitating entangled photons.
As an example, consider a massive unstable particle coincident with the infalling geometrically
stationary system that decays into an entangled pair of photons just as the
particle encounters the outer trapping surface $R_S (ct_{decay})$.  One of the
photons will be emitted radially outward, and the other radially inward to conserve
microscopic momentum.  One observer (Alice) will be a geometrically stationary
observer with final location $x=R_{So}$, while the other observer (Bob)
will be a geometrically stationary observer located diametrically opposite
the original observer with final location $x=-R_{So}$.  

This arrangement is
demonstrated in Figure \ref{SpTmEntangle}.  The left diagram of the figure is a spacetime plot using the coordinates $(ct,x)$, while the right diagram is a conformal plot derived from $(ct,r)$.
\twofigures{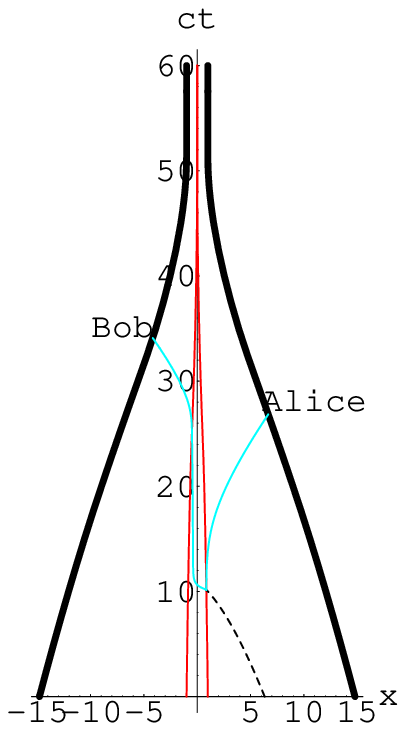}{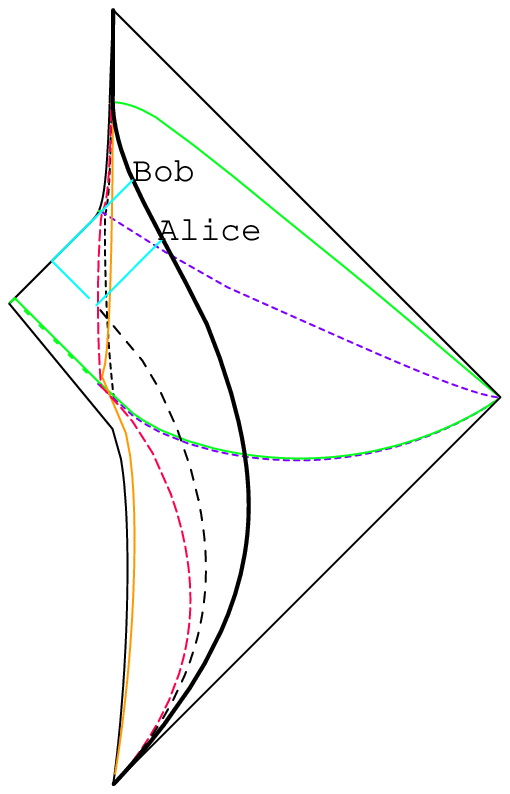}{Trajectories of the entangled photon pair
emitted by the unstable particle just as it crosses into the trapped region.  The left
diagram is a standard spacetime diagram using $(x,ct)$, while the right diagram
utilizes conformal (radial) coordinates $(Y_\rightarrow , Y_\uparrow)$ and superposes two opposite values of the azimuthal angle $\phi$.}
The infalling unstable particle is represented by the dashed black curve, while
the entangled photons which are the products of the decay are represented
as the outgoing and ingoing null light-blue trajectories, ultimately
detected by Alice and Bob.  In the spacetime diagram on the left, the
interior mass scale $R_S (ct)$, which coincides with the outer trapping
surface while there is a black object present, is also demonstrated
as a pair of solid (red) timelike curves slowly approaching zero.  

The decay occurs just as the unstable particle encounters the trapping surface,
indicated by the initial vertical slope of the outgoing photon trajectory
that ultimately reaches Alice. 
The ingoing photon crosses the center, but remains trapped until the
trapped region vanishes due to the evaporation of the black object, at which time
it crosses the interior mass scale and ultimately is detected by Bob. 
In the radial conformal diagram on the right, Alice has azimuthal location
$\phi=0$, while Bob has azimuthal location $\phi=\pi$.  Thus, these observations are actually
spacelike separated, which is not depicted on the radially symmetric
Penrose plot. In this diagram, the ingoing entangled photon ``reflects"
from the center as its polar angle switches from zero to $\pi$.
It is interesting to note that, in contrast to the situation in flat spacetime,
Alice and Bob likely detect the photons during differing epochs.  In
addition, the energy ratio $\epsilon \equiv {u_\gamma ^{ct} \over u_{\gamma*}^{ct}}$ measured by Alice
is $\epsilon \approx 0.01$, while that measured by Bob is $\epsilon \approx
8 \times 10^{-8}$.  Thus, the spacetime
and energy-momentum entanglement information
measured by the different observers is both
temporally-shifted and redshifted in energies.

\setcounter{equation}{0}
\section{Conclusions}
\indent

A singularity-free transient black object whose center remains always timelike
has been developed, and its evolution has been numerically explored. 
Such objects are difficult to distinguish from long-lived transient black
holes in the exterior, yet are everywhere analytic.
The geometry was constructed to satisfy energy conditions everywhere
during accretion.  
In addition, the geometry satisfies energy conditions external to the trapped region, and external to any region expected to involve prolific production of quanta.  
The coordinates utilized are particularly useful since they admit
a class of geometrically stationary (co-moving) observers.  These observers
share proper time with that of the asymptotic observer, directly
parameterized by the metric time $t$.  Geometrically stationary
observers can serve as convenient platforms for performing
measurements within the dynamic geometry.

The particular model explored involves a geometry with an overall
energy distribution of fixed mass $M$ that undergoes a pressureless collapse initiated
in the distant past.  The collapse terminates after quantum non-locality
effects are presumed to dominate the dynamics \cite{JL2011}.  Once mass is contained within
a region smaller than its Schwarzschild radius,
a trapped region ($i.e$ a region in which all causal trajectories move toward
decreasing radial coordinate) forms.  However, since the center remains
timelike and the black object is transient, no horizon develops, and the object never becomes a black hole.  This collapsed black
object then undergoes thermal decay analogous to that
expected from a black hole until the trapped region vanishes.  The
remnant of thermal evaporation is then chosen to decay away in a manner consistent
with quantum and geometric constraints.

A conformal diagram demonstrating the large-scale causal structure of
the geometry has been demonstrated.  The modification of the
spacetime from that of Minkowski is minimal, since the center $r=0$ remains
everywhere timelike.  In these coordinates, the development of a trapped region merely
expands the volume of the conformal diagram, as expected from previous studies \cite{JLTF2010, BABJL3}. 
Fixed-time surfaces remain everywhere spacelike, providing global foliation
for parameterizing the dynamics.  Surfaces of fixed radial coordinate
have been demonstrated to be timelike exterior to the trapped region, and spacelike within the trapped region.

Information exchange in the vicinity of the black object has also been explored. 
Outgoing photons from an infalling
emitter were seen to redshift as the emitter approaches the trapped region.  However, in contrast
to a transient black hole, direct emissions that are temporarily trapped while
the black object persists are later observed once the trapped region vanishes. 
This rapid release of information has been demonstrated for the model. 
The redshift of energies and the dynamics of the rate of communications
are shown to behave as expected.

In order to explore the dynamics of entangled information, the trajectories of
entangled photons have been examined.  The photons were formed as a particle decayed while crossing into the trapped region.  The presence of the transient black object alters
the relative times of observation and energy redshifts of the entangled
photons.  The exploration directly demonstrates that the loss of entanglement
information is only temporary for this geometry, and that there are no obvious cases of violation of the standard laws of physics. 

The analytic and causal properties
of this dynamic black object should considerably simplify the exploration of quantum
geometrodynamic behaviors on the geometry, as will be demonstrated elsewhere \cite{JLCUPress}. 
Also, one should be able to incorporate such a transient black object within
a dynamic de Sitter geometry consistent with big bang cosmology, as has been done
with a transient black hole \cite{JLTF2010}.  In principle, such a transient
black object should introduce no exotic behaviors in the global geometry, while yet
introducing an additional trapped region into the spacetime.

\bigskip
\begin{center}
\textbf{Acknowledgments}
\end{center}

The authors warmly acknowledge their association with
Beth Brown.  TF is pleased to acknowledge past discussions with Ted Jacobson, along with support from Chanda Prescod-Weinstein.  JL gratefully acknowledges useful past discussions with 
James Bjorken, Paul Sheldon, and Lenny Susskind.

\end{document}